\documentclass[prb,preprint]{revtex4-1} 

\usepackage{amsmath}
\usepackage{amsfonts}
\usepackage{graphicx}

\usepackage[hidelinks]{hyperref}
\newcommand{\orcid}[1]{\href{https://orcid.org/#1}{\includegraphics[width=11pt]{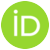}}}
\usepackage[dvipsnames]{xcolor}

\begin{document}

\title{Visualization of cylindrical resonances}

\author{Brais Vila\orcid{0000-0001-8124-6616}}
\email[Electronic mail: ]{boubina@alumni.unav.es}
\affiliation{Focke Meler Gluing Solutions, S.A. (Arazuri, Spain)}

\date{August 13, 2023}

\begin{abstract}
The analysis of cylindrical resonators is part of standard physics curricula but, unlike for their rectangular counterpart, their mode structure is hardly ever visualized. The aim of this work is to show a way of doing it, providing a set of interactive web applications and citing potential use cases in the form of both academic courses and published research. These cover several branches of physics and engineering, showing that these materials can be useful for a broad audience.
\end{abstract}

\centerline{\large{\textbf{Visualization of cylindrical resonances}}}
\vspace{1cm}
This version of the article is the accepted manuscript, after peer review and before copyediting for the \textit{European Journal of Physics}. IOP Publishing Ltd is not responsible for any errors or omissions in this version of the manuscript or any version derived from it. The Version of Record is available online at \textcolor{blue}{\url{https://doi.org/10.1088/1361-6404/acf5b6}}.
\subsection*{Supplementary material}
This article makes reference to a set on online supplements. A copy is hosted by IOP publishing at the URL cited above, but the supplements are not part of the Version of Record. Another copy has been uploaded as \href{https://arxiv.org/src/2310.18514v2/anc/Index.html}{\textcolor{blue}{ancillary files}} to the record \href{https://arxiv.org/abs/2310.18514}{\textcolor{blue}{arXiv:2310.18514}}.
\subsection*{Copyright}
The \href{https://www.eps.org}{\textcolor{blue}{European Physical Society}} holds copyright to the Version of Record. More information can be found on the \href{https://crossmark.crossref.org/dialog/?doi=10.1088/1361-6404/acf5b6&domain=pdf&date_stamp=2023-10-09}{\textcolor{blue}{CrossMark entry}} of the article.

The present version of the manuscript is shared by the author under a \href{https://creativecommons.org/licenses/by-nc-nd/4.0/}{\textcolor{blue}{CC BY-NC-ND}} license. It is done after a 12-month embargo during which the \href{https://arxiv.org/abs/2310.18514v1}{\textcolor{blue}{version before peer review}} was shared.
\subsection*{IOP Publishing policies}
More information about \href{https://publishingsupport.iopscience.iop.org/preprint-pre-publication-policy}{\textcolor{blue}{Preprint policy}}, \href{https://publishingsupport.iopscience.iop.org/publishing-support/authors/copyright-and-permissions}{\textcolor{blue}{Copyright and Permissions}} and sharing of \href{https://publishingsupport.iopscience.iop.org/questions/supplementary-material-and-data-in-journal-articles}{\textcolor{blue}{Supplementary materials}} are published on the website of IOP Publishing.
\subsection*{Broken-link policy}
This document will not be updated, even if any of the links above becomes inactive. However, the author is available to provide further information or assistance upon request.
\newpage

\maketitle

\section{Introduction}\label{intro}
Cavity resonators are studied in both undergraduate\cite{labSantaB} and graduate\cite{uspas} courses because they are ubiquitous: from their unexpected effects in the manufacturing of industrial goods\cite{witting} to their complex engineering for quantum computing\cite{choiT} and nuclear fusion\cite{thumm}, the likelihood that physics and engineering students will find them in their future careers is very high.

Ease of manufacturing favors rectangular and cylindrical cavities, but the latter offer the higher quality factor\cite{jackson} and are therefore the single most relevant type in practice, even more so when rectangular shapes are impractical due to size\cite{hattori} or curvature-dependent dynamics\cite{franke}.

Laboratory assignments involving these devices are common in undergraduate courses because they offer an accessible and affordable opportunity to work on relatively advanced topics \cite{labSantaB,ffield,elias,moloney,labPurdue,barreiro,oliveira,jaafar,varberg}. For instance, they can be used to measure the density of a plasma\cite{ffield,oliveira}, visualize sound waves\cite{elias,jaafar}, demonstrate the photoacoustic effect\cite{barreiro}, quantify fundamental constants\cite{labSantaB,labPurdue} and determine gas properties like the speed of sound\cite{labSantaB,labPurdue,varberg}, molecular mass \cite{labSantaB} and heat-capacity ratio\cite{varberg}. These assignments usually deploy techniques not covered in theory classes, leaving a gap that can be closed using the web applications distributed with this article\cite{supplements}. The main idea is that those concepts can be rapidly understood with adequate visualizations of the most important formulas in the theory, a method commonly used to study wave phenomena.\cite{dori,murello,girwidz,franklin,ghali,ahmed,chhabra}. Therefore, this article contains almost no mathematical formulas, favoring screenshots of the applications instead. However, all the underlying mathematical details are explained in two separate files named \textit{documentation} and \textit{mathematical model}\cite{supplements}, giving the reader every opportunity to learn the complete theory.

The starting point for the applications is a very simple idea, presented in section \ref{structure}, about how to visualize the information contained in the formula for the resonant frequencies of a cylindrical cavity. Section \ref{field} explains why it is important to represent the mode structure along with the field distribution associated with the resonant modes: it helps to understand which mode should be used for a specific application. Some general aspects about how that can be taken to practice are discussed in section \ref{coupling}. Since these are all aspects concerning the design process, its importance is examined in section \ref{design}. Section \ref{discussion} explores to which extent those analyses can be applied to other types of problems and the conclusions are summarized in section \ref{conclusions}.

\section{Visualizing the mode distribution}\label{structure}

The key concepts a student must understand to undertake some laboratory assignments\cite{ffield,labPurdue,jaafar,varberg} are related to the existence and frequency distribution of the resonant modes, rather than their individual characteristics. For instance, knowing that the frequency of the modes is directly proportional to their index and to the speed of sound suffices to determine the latter measuring the former\cite{varberg}. To understand that, let us have a look at Fig.~\ref{longMode}, which is an annotated screenshot of Supplement A. Here, the waves travelling from both sides have collided at the center and are about to be reflected back. When they reach the ends of the tube, pressure will accumulate there and the direction of propagation will shift again, completing the oscillation when the situation in Fig.~\ref{longMode} is repeated. Therefore, the speed of sound is $c=2(l/2)f$, where $l$ is the length of the tube and $f$ the frequency of oscillation. More generally, what Supplement A shows is that for every $n_z\in\mathbb{Z^+}$ there is a resonant mode verifying $2lf_{n_z}=cn_z$. Measuring $f_{n_z}$ and plotting it against $n_z$ lets us calculate $c$ from the slope\cite{varberg}.

\begin{figure}[h!]
\centering
\includegraphics[scale=1]{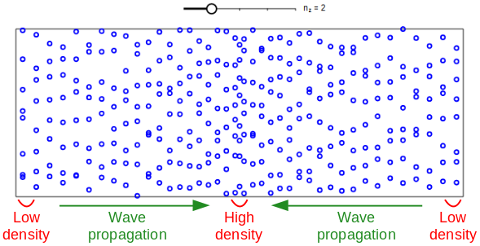}
\caption{Second longitudinal mode of a cylindrical resonator (supplement A).}
\label{longMode}
\end{figure}

The geometric argument can be extended to more complex cases\cite{redwood} but the more general approach is to solve the wave equation with adequate boundary conditions\cite{redwood,balanis,morse}, leading to Eq.~\eqref{resFreq} for a closed cylindrical resonator of length $l$ and radius $a$:
\begin{equation}
\label{resFreq}
f_{m,n,n_z}=\frac{c}{2}\sqrt{\left(\frac{n_z}{l}\right)^2+\left(\frac{\alpha_{mn}}{a}\right)^2} \hspace{2em} \begin{array}{l} m=0,1,2... \\ n_z=0,1,2...\end{array}
\end{equation}

Here, $\alpha_{mn}=x'_{mn}/\pi$, with $x'_{mn}$ being the $n^\text{th}$ zero of $J'_m(x)$, the first derivative of the Bessel function of the first kind and order $m$. Eq.~\eqref{resFreq} results from imposing the boundary conditions for a completely closed resonator on the solutions of either the scalar or vector wave equation. For instance, the electric field verifies Eq.~\eqref{waveEq} and must be normal to the boundaries, which is true if the latter are perfect conductors and can be written $\hat{n}\times\vec{E}=0$. For an acoustic pressure field, $p$ would replace $\vec{E}$ in Eq.~\eqref{waveEq} and the velocity of the particles cannot be normal to the boundary, which can be written as $\hat{n}\cdot\vec{\nabla}p=0$.

\begin{equation}\label{waveEq}
\nabla^2\vec{E}-\frac{1}{c^2}\frac{\partial^2\vec{E}}{\partial t^2}=0
\end{equation}

For acoustic resonances, it is customary\cite{morse} to index the zeros $n=0, 1, 2...$ whereas for electromagnetic resonances $n=1, 2, 3...$ is used\cite{balanis}. Although that is ultimately a matter of notation, there is a clear reason supporting that convention, which is related to whether the trivial solution should be included as a root of $J'_m(x)=0$ and is explained in the documentation of the supplements\cite{supplements}. For a similar reason, $n_z=0$ must be excluded for transverse electric (hereafter, TE) modes. For transverse magnetic (hereafter, TM) modes, the zeros of $J_m(x)$ appear instead of those of $J'_m(x)$.

\begin{figure}[h!]
\centering
\includegraphics[scale=0.5]{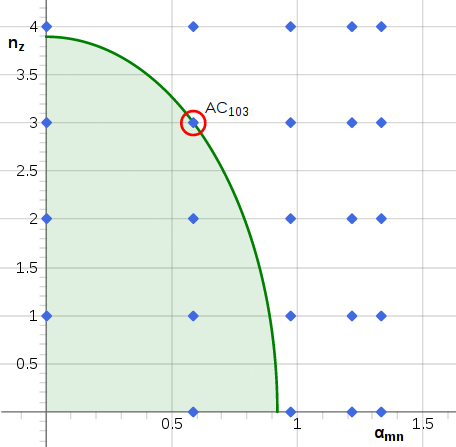}
\caption{Mode distribution of a cylindrical acoustic resonator (supplement B).}
\label{modeDist}
\end{figure}

The fact that the modes are evanescent below their resonant frequency\cite{redwood} is represented by the inequality $f\geq f_{m,n,n_z}$. Along with Eq.~(\ref{resFreq}), it represents the region under an ellipse with semi-axes $2fa/c$ and $2fl/c$ on the $\alpha_{mn},n_z$-plane. This suggests drawing a plot like Fig.~\ref{modeDist}, where each mode is represented by a point and those that can actually resonate are contained in the green-shaded area. From them, those near the border (the solid green curve) are more likely to be excited, but ultimately that depends on the feed. A general description of how that happens is provided in section~\ref{coupling}.

Supplement B generates interactive plots like Fig.~\ref{modeDist}: the user can vary the parameters using the sliders on top and see how the ellipse changes. In Fig.~\ref{modeDist}, seven modes can potentially be excited but only $AC_{103}$ resonates at the selected frequency. The number of modes below the curve can be counted using supplement C. Supplements D and E do the same for cylindrical electromagnetic resonators.

It is surprising that visualizing the mode structure, as shown in Fig.~\ref{modeDist}, is hardly ever attempted for cylindrical cavities. Remarkable exceptions come as three-dimensional plots with a complex mode distribution\cite{morse} or even more intricate two-dimensional charts\cite{nyfors,mehdizadeh}. The plots presented here are simpler and easier to understand. In contrast, this kind of visualization is commonplace for rectangular resonances. An example of this kind of mode counting from undergraduate solid-state physics courses is the density of states of phonons in a crystal\cite{kittel}. The exact same approach is used when counting modes in a rectangular microwave cavity\cite{meredith} or a rectangular acoustic chamber\cite{morse}.

Not visualizing the mode structure leads to missing an important piece of information: the very reason why the plot in Fig.~\ref{modeDist} works is that the zeros of Bessel functions are ordered\cite{palmai}. The relevant interlacing properties are summarized at the end of the documentation of the supplements\cite{supplements}. Students are rarely given that information, without which the frequencies in Eq.~\eqref{resFreq} cannot be ordered.

\section{Visualizing the field distribution}\label{field}

\begin{figure}[h!]
\centering
\includegraphics[scale=0.3]{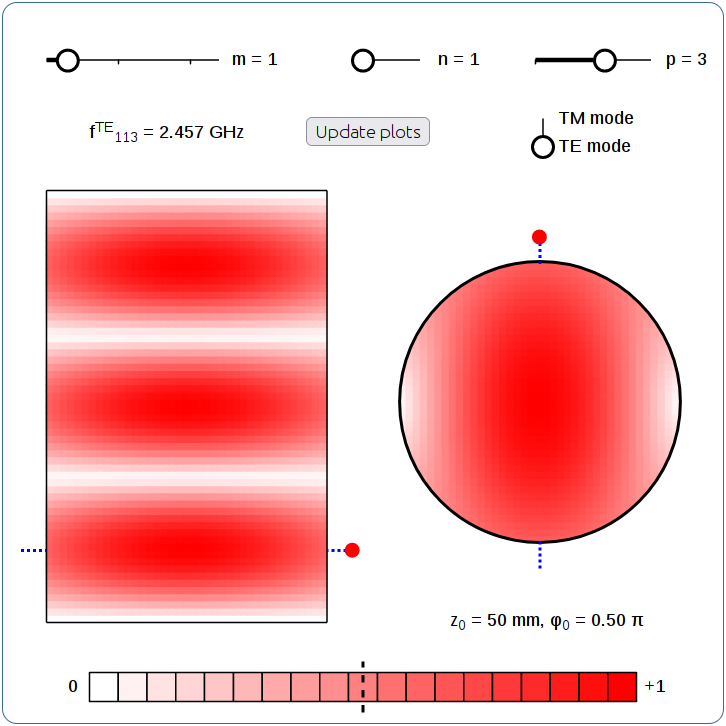}
\caption{Electric field amplitude of the $TE_{113}$ mode (supplement D).}
\label{fieldDist}
\end{figure}

Some laboratory assignments require that the students have a good understanding of the characteristics of specific modes\cite{barreiro,labSantaB,elias,moloney,oliveira}. For instance, plasma-density measurements can be carried out aligning two loop antennas with the electric field maxima of the $TM_{010}$ mode\cite{oliveira}. The same is true for light sources in photoacoustic excitation\cite{barreiro}. Experiments to visualize the modes themselves have also been proposed\cite{elias}. Those cases can be explained using plots like Fig.~\ref{fieldDist}, which are usually obtained by running a simulation and included in publications treating resonators\cite{choiT}. The web applications accompanying this article offer a more portable and lightweight solution: in supplements B and D, the user can navigate through the modes and see their shape using the applet on the right. The coordinates of these plots can be chosen dragging the red dots.

The mode in Fig.~\ref{fieldDist} can be used to design an electron cyclotron resonance (hereafter, ECR) ion source or an X-ray source, but the presence of two (white) nodal lines at $z=l/3$ and $z=2l/3$ requires adjusting the magnetostatic field accordingly, which can only be avoided in the $TE_{111}$ mode\cite{orozco}. The fact that we can visualize this information while we choose the height and radius of the cavity and see where the mode is located in relation to the ellipse from Fig.~\ref{modeDist} is extremely practical and, in the author's opinion, should be implemented in professional simulation packages commonly used for designing resonators.

Visualizing the modes is important because having a good qualitative grasp of their properties is necessary. For instance, $TM_{mn0}$ modes are useful because their fields are longitudinally uniform\cite{choiT,oliveira}. The multi-cell coupled resonators used in particle accelerators\cite{gerigk} are a remarkable example. The $TM_{010}$ mode resonates in each cell to reduce the difference between phase and particle velocities\cite{uspas}. TM modes of higher longitudinal order present a particularity that is normally not mentioned in the literature. The reader is encouraged to open supplement D and plot the $TM_{111}$ mode for a cavity with dimensions $a\approx 5~mm$, $h\approx 100~mm$. After changing the radius to $a\approx 350~mm$, a warning appears that the plots are not updated. Refreshing the plot renders a completely different field distribution. This is a peculiarity of TM modes: no warning is issued when doing the same for TE modes because their field distribution never changes (and neither does that of the acoustic modes in supplement B and the quantum-mechanical modes in supplement F). The explanation for this nuance is given in the documentation of the supplements along with other mathematical details\cite{supplements}.

\section{Understanding coupling}\label{coupling}

\begin{figure}[h!]
\centering
\includegraphics[scale=0.44]{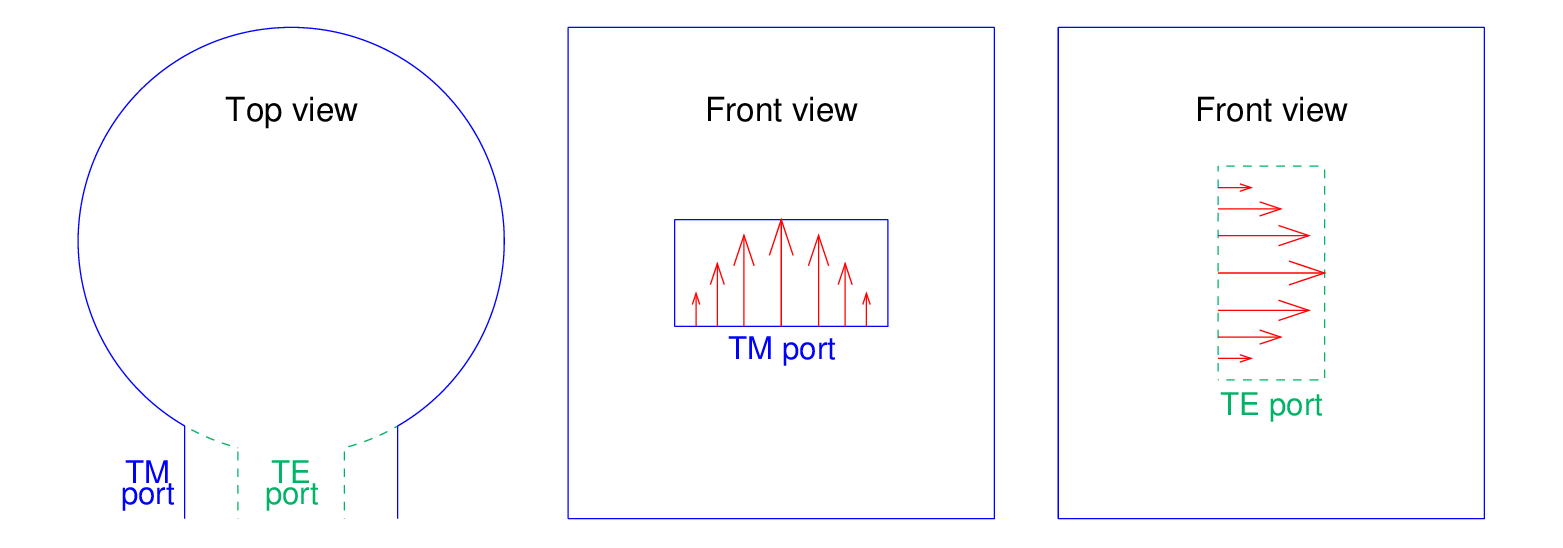}
\caption{Rectangular waveguide feeding a cylindrical resonator.}
\label{figEM}
\end{figure}

Practical resonators must have some kind of input port and they might have an output port or a load inside. For the analysis presented in sections \ref{intro} to \ref{field} to be valid, there must exist a coupling mechanism between those other elements and the resonant modes of the cavity. For a simple example, let us revisit Fig.~\ref{fieldDist} and think how that mode could be excited using a rectangular waveguide, as shown in Fig.~\ref{figEM}. Assuming a $TE_{01}$-mode electric field in the waveguide, indicated by the red arrows, it is obvious why the orientation shown on the right-hand side is the better option. However, the reader must be aware that it might be possible to excite a TE mode with a TM port, particularly if the field at the junction is zero. Supplement D can be used to visualize why that is more likely to happen with modes $TE_{112}$ or $TE_{114}$ than in Fig.~\ref{fieldDist}. More complex feeds for specific modes are described in the literature\cite{balanis,mehdizadeh,meredith}. 

\begin{figure}[h!]
\centering
\includegraphics[scale=0.4]{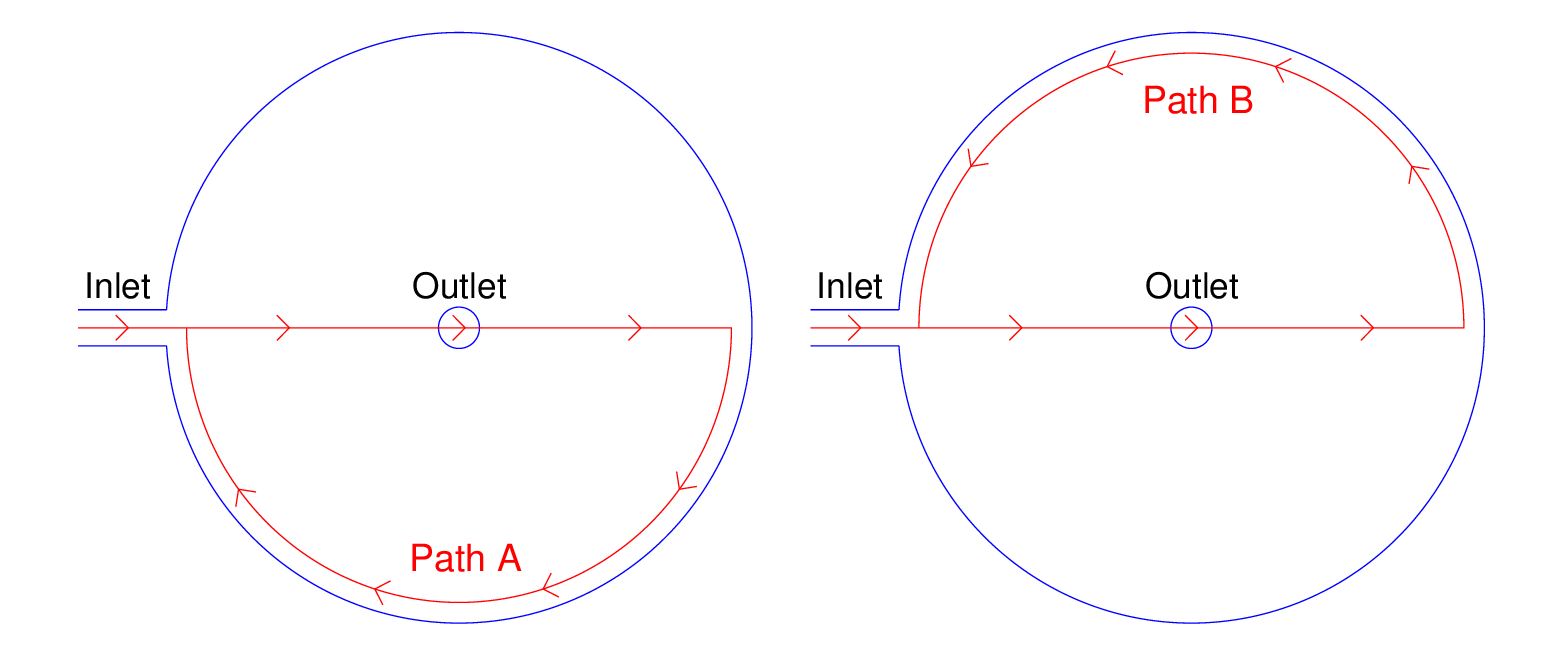}
\caption{Proposed dynamics for the experiments described in Ref.~\onlinecite{franke}, where the jet exiting the outlet moves sideways as seen from the inlet, oscillating at\cite{franke} 990 Hz. The analytical value calculated with supplement B is 988 Hz.}
\label{figAcoustics}
\end{figure}

An analogous situation takes place when an acoustic source is used\cite{labPurdue,labSantaB,elias,moloney,jaafar} and coupling occurs in a similar manner\cite{redwood}. The analysis differs a little when the excitation is not of acoustic nature yet has a well-defined frequency\cite{hattori,witting,barreiro}, but it can still get trickier: acoustic resonances can also be triggered by fluid flows that are not inherently oscillatory without the presence of the cavity\cite{franke}. To study how coupling might still occur, let us think of a cylindrical cavity with a radial inlet on the curved wall and an axial outlet on one of the flat walls, as shown in Fig.~\ref{figAcoustics}.

When the inlet is kept at constant pressure, the jet exiting the outlet oscillates at the resonant frequencies given by Eq.~\eqref{resFreq}. These oscillations have been observed experimentally\cite{franke}, but a theoretical explanation has not been given. To fill that void, a \textit{toy model} is included in the supplemental material accompanying this article, accessible through the link \textit{mathematical model}\cite{supplements}. It is based on the assumption that the jet switches between the two paths shown in Fig.~\ref{figAcoustics}, coupling with the fundamental mode of the cavity. The latter can be visualized using the right-hand side of supplement B to plot the mode $AC_{100}$. For the left-hand side, parameters $a\approx 100~mm$, $l\approx 80~mm$ and $f\approx 1~kHz$ should be used. 

\begin{figure}[h!]
\centering
\includegraphics[scale=0.56]{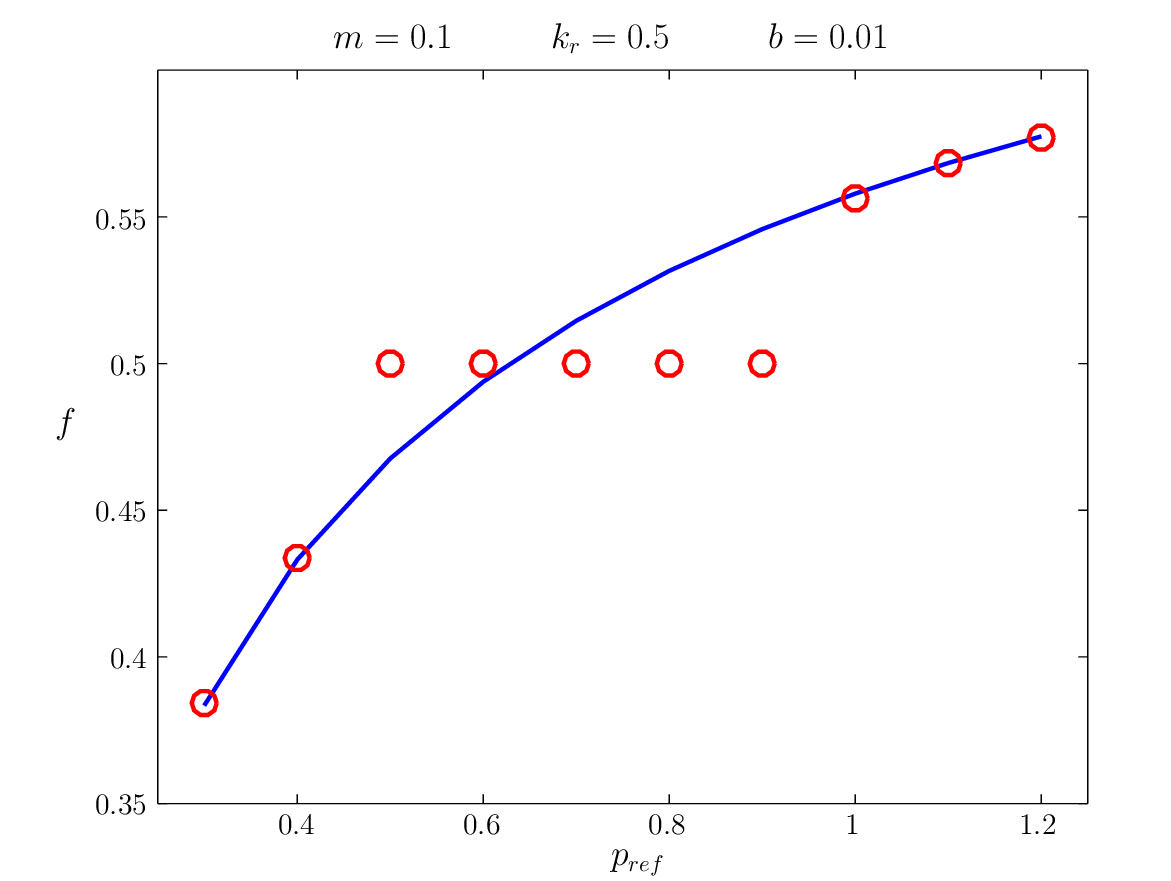}
\caption{Frequency response of the model associated with Fig.~\ref{figAcoustics}. The line shows the behavior without the presence of resonant modes and the circles with a resonant mode of dimensionless frequency 0.5.}
\label{figModel}
\end{figure}

Despite its simplicity, the model gives the reader a sound understanding of what coupling is and reproduces many of the general aspects of cavity resonators. Most notably, it shows how self-sustained oscillations of the jet due to its interaction with the cavity couple with its first resonant mode, resulting in the frequency response shown in Fig.~\ref{figModel}: for a certain range of inlet pressures, coupling leads to very stable operation.
More details can be found in the supplemental material\cite{supplements}, but it is worth noting that the participating phenomena are very distinct. In principle, the oscillations arise from the interaction between a jet that can move at different velocities (either subsonic or hypersonic) and a cavity. The faster the jet, the higher the frequency of oscillation. The size of the cavity only matters relative to that velocity: if the size and velocity are both reduced, the frequency might remain the same. Consequently, fluidic oscillators can be scaled up or down significantly and maintain their operation. More formally, we say that these devices operate at approximately constant Strouhal number. This is a problem of fluid dynamics, in which the fluid particles travel long distances. On the other hand, as shown in Supplement A and Fig.~\ref{figAcoustics}, the particles that propagate a resonant wave do not travel, but simply oscillate. This is a problem of acoustics, in which the wavefront always moves at the speed of sound. It is also not scalable: since the velocity is always the same, the size determines the frequency of oscillation, given by Eq.~\eqref{resFreq}. The latter is an optional input of the model: without it, the model yields a solution to the fluid dynamics problem; with it, one to the coupled problem.

This kind of complex interaction is common in cavity resonators: coupling might occur between modes of the feed and the cavity\cite{maksimov}, a cavity and a load\cite{elnaggar}, an electron beam and a cavity\cite{kumar,orozco}, an airflow and a cavity\cite{franke} and many other pairs. No general method will be valid to analyze every case, but a good understanding of the modes and their distribution is a prerequisite for every approach.

\section{Designing the resonator}\label{design}
In sections \ref{field} and \ref{coupling}, we have seen two examples of how a resonator might be used in the form of an ECR ion source and a fluidic oscillator. For the latter, a theoretical model of the coupling between its resonant modes and the incoming air jet has given us Fig.~\ref{figModel}, which shows that the resonator can be used to stabilize the oscillator. The key word here is \textit{theoretical}: what if the pressure range for which coupling occurs is so low that we cannot get the flow rate we need? What if it is so high that turbulence makes it useless? Similar questions can be asked about the ion source: we need a TE mode to interact with the helical motion of the electron, but what if the ohmic loss on the walls of the cavity is too high for the $TE_{111}$ mode? We might then choose a higher $TE_{11p}$ mode, but that makes monomode excitation harder to achieve and, as mentioned in section \ref{field}, requires adjusting the magnetostatic field accordingly. Will we actually be able to manage a practical solution?

In this sense, we could describe \textit{design} as the process that takes us from what is theoretically conceivable to what is practically achievable with a resonator. Fortunately, both of the aforementioned examples do actually work and a plethora of simple uses only require straightforward designs, but physicists and engineers are currently tackling extremely difficult technical problems designing resonators in areas like quantum computing\cite{choiT} and nuclear fusion\cite{kumar,thumm}.

Our first decision when designing a resonator will be opting for either a \textit{monomode} or a \textit{multimode} cavity. By now, the reader should be able to visualize in Fig.~\ref{modeDist} that, strictly speaking, only two types of monomode cavities exist ($AC_{001}$ and $AC_{100}$), but a proper design will allow us to select other modes like $AC_{103}$. Supplement D can be used to visualize the electromagnetic case.

In general, monomode cavities are used whenever the field distribution needs to be known. This includes low-power applications like sensing and measurement, but the fact that monomode cavities can be designed to concentrate high fields in specific locations also makes them perfect for power applications like sintering, plasma ignition, maser and laser excitation and telecommunications devices such as antennas and filters.

\begin{figure}[h!]
\centering
\includegraphics[scale=0.46]{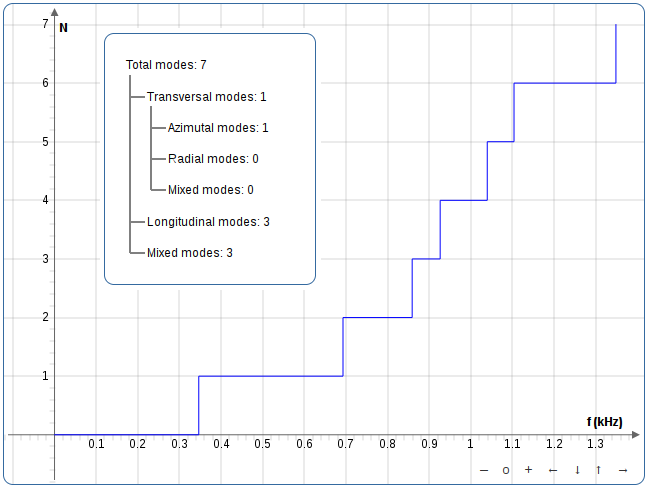}
\caption{Number of modes contained in the green-shaded area in Fig.~\ref{modeDist} (supplement C).}
\label{figCount}
\end{figure}

Multimode cavities are studied for room acoustics and microwave processing of large loads. The most popular example is the household microwave oven, which can also be used in the undergraduate laboratory\cite{barnes}. They are designed to have as many modes as possible and mode-counting plots for rectangular cavities are present in the literature\cite{morse,meredith}. Analogous plots for cylindrical cavities, like Fig.~\ref{figCount}, are equally important but much harder to find. They can be generated using supplements C and E. The latter can also be used to illustrate how difficult monomode coupling can be in advanced applications like nuclear fusion\cite{kumar}: introducing $a=22.6~mm$, $h=0$ and $f=(240\pm 0.5\%)~GHz$ yields 1621 and 1586 modes respectively, for a cumbersome total of 35 competing modes within that range.

The design process varies greatly from one application to another and between industry and academia. For instance, a chemist might want to gently heat a liquid in a small plastic test tube to enhance a chemical reaction. A microwave designer contracted for the task could open supplement D, introduce the parameters $a\approx 100~mm$, $h\approx 45~mm$ and visualize the mode $TM_{210}$, which can be excited with the TM port shown in Fig.~\ref{figEM}. The plot shows that the field vanishes at the curved wall, which favors two electric field maxima being aligned with the waveguide longitudinal axis. Like the outlet in Fig.~\ref{figAcoustics}, a small pipe can be welded on top of each maximum to process four samples, which would only penetrate a few millimeters into the cavity to prevent excessive microwave absorption leading to boiling. Such a small perturbation barely affects the fields, which would be confirmed via simulation. That would be a common industrial procedure that, for a simple case like this, would be completed in a couple of days. In contrast, it took 17 years of innovative scientific work to advance from using a $TE_{0,2}$ mode to $TE_{25,10}$ in the cavities of gyrotrons for fusion plasma heating\cite{thumm}.

\section{Modifying the applications}\label{discussion}

The source code of the applications accompanying this article is made available to the public without restrictions of use. The reader can therefore tweak them at will, implementing minor adjustments like changing the limits of the sliders. More-significant changes can be made to visualize other types of problems. For instance, if we were studying the creation of photons in a cylindrical resonator with time-dependent length, we could note that the parametric resonant case, in which the number of photons grows exponentially, takes place when a specific mode is located exactly halfway between the origin and the ellipse\cite{crocce}. Supplement D could be modified to show that explicitly. Another example is the \textit{particle-in-a-box} model, taught in the vast majority of undergraduate courses in quantum mechanics. The cylindrical case is included in standard textbooks\cite{liboff} and has been used to study electron behavior in nanowires and graphene sheets\cite{baltenkov}. Its analytical solutions have been implemented in supplement F, where the probability of a particle being at a certain position inside the cylinder, or having its momentum pointing in a specific direction, can be visualized. The left-hand side of the application, analogous to Fig.~\ref{modeDist}, raises two natural questions. First, whether it makes sense to speak of a well-defined energy (represented by the ellipse) for a particle in a superposition of states. Second, whether only the states with lower energy values (those below the ellipse) can actually be detected. The answer to both questions is no\cite{rohrlich} and the ellipse is drawn here just as a reference for the energy levels of the modes. The fact that all these questions come up naturally makes supplement F worth including, especially considering that it has been shown that students tend to have difficulty understanding the meaning of energy eigenvalues \cite{chhabra}. Yet another question comes up naturally: if acoustic resonances were described in terms of particle movement in section \ref{structure}, with surface currents playing the same role in the electromagnetic case, is it possible to find a similarly intuitive interpretation of this quantum resonance? A very interesting answer is considering the Schr\"odinger equation a diffusion equation and describing the modes in terms of probability currents, although such an interpretation has only been shown possible in the one-dimensional case and is far from trivial to scale up to higher dimensions\cite{mita}.

\section{Conclusions}\label{conclusions}
We can visualize the mode structure of a cylindrical resonant cavity using very simple two-dimensional graphs that, presented interactively and combined with plots of the relevant fields, serve a double purpose: on the one hand, they can be a great aid for students learning about resonators; on the other, they can be used as a design tool in an industrial environment.

That idea has been implemented in a set of JavaScript applications and this paper provides a gradual explanation of how they can potentially be used: sections \ref{structure} and \ref{field} are adequate for undergraduate students in preparation for certain laboratory assignments, sections \ref{coupling} and \ref{design} offer an accessible introduction to topics that graduate students will encounter working with resonators and, lastly, section \ref{discussion} touches upon potential uses of the applications beyond those for which they were originally designed.

While resonant cavities are part of standard physics curricula, the time allocated in theory courses is usually insufficient to cover all aspects relevant for laboratory sessions. The main contribution of this article is providing a visual manner of teaching the theory, under the premise that it will speed up concept acquisition and allow for more-complete preparation of laboratory work. However, starting with an intuitive approach to the topic does not mean that the learning process must end there: all the mathematical details about how the applications work are explained in the accompanying documentation.
\begin{acknowledgments}

The author wishes to thank the Government of Navarre for partially funding this work under the programme \textit{Industrial doctorates 2018-2020}.

\end{acknowledgments}

\section*{Author Declarations}
The author has no conflicts to disclose.

\end{document}